\documentstyle[editedvolume,psfig]{crckapb} 

\newcommand{\gtilde}
 {~ \raisebox{-1ex}{$\stackrel{\textstyle >}{\sim}$} ~}

\begin{opening}
\title{Luminosity Density Evolution in the Universe and
       Cosmological Parameters}

\author{Tomonori Totani}
\institute{Department of Physics, The University of Tokyo \\
           Tokyo 113, Japan \\
           totani@utaphp2.phys.s.u-tokyo.ac.jp}

\end{opening}

\runningtitle{LUMINOSITY DENSITY EVOLUTION IN THE UNIVERSE}

\begin{document}

\begin{abstract}
Star formation history in galaxies is strongly correlated to
their present-day colors and the Hubble sequence can be considered
as a sequence of different star formation history.
Therefore we can model the cosmic star formation history based on
the colors of local galaxies, and comparison to direct observations
of luminosity density evolution at high redshift gives a new test for
the cosmological parameters which is insensitive to merger history of 
galaxies. The luminosity density evolution in $0<z<1$ observed by the
Canada-France Redshift Survey in three wavebands of 2800{\AA}, 4400{\AA}, 
and 1$\mu$m indicates that the $\Lambda$-dominated flat universe
with $\lambda_0\sim0.8$ ($>$ 0.53 at 95\%CL) is strongly favored.

The cosmic star formation rate (SFR) at $z>2$ is also compared to the
latest data of the Hubble Deep Field including new data which were
not incorporated in the previous work of Totani, Yoshii, \& Sato (1997), 
and our model of the luminosity density
of spiral galaxies taking account of gas infall is consistent
with the observations. Starbursts in elliptical galaxies, which are
expected from the galactic wind model, however overproduce SFRs and hence 
they should be formed at $z \gtilde 5$ or their UV emission has to be
hidden by dust extinction. The amound of metals in galactic winds
and escaping ionizing photons are enough to contaminate the Ly$\alpha$
forests or to reionize the universe.
\end{abstract}

\section{Introduction}
Recently the cosmic star formation history is becoming a very
hot topic mainly because direct observations of luminosity density
evolution in the universe at high redshifts become possible.
The cosmic star formation history is itself very interesting,
but here we stress that it is also important for many related
topics in astrophysics, e.g., relic supernova
neutrinos (Totani, Sato, \& Yoshii 1996), estimate of the cosmological
parameters (Totani, Yoshii, \& Sato 1997; TSY), and gamma-ray bursts
brightness distribution analysis (Totani 1997). Here we describe
mainly the work of TSY which showed that the marked evolution
in the luminosity density of galaxies
in $z<1$ observed by the Canada-France Redshift Survey (CFRS;
Lilly et al. 1996) favors the existence of non-vanishing cosmological
constant, $\Lambda$, in all the three wavebands of restframe 2800{\AA}, 
4400{\AA}, and 1$\mu$m. 
The latest data of cosmic star formation rate
at $z>1$, which were not available in TYS, are also newly incorporated here
and star formation history at higher redshifts is discussed.

The present-day colors of galaxies are known to be 
strongly correlated with their morphological type through
type-dependent history of star formation, and therefore it is possible
to construct a model of cosmic star formation history by summing up
type-dependent evolution models of galaxies with relative proportion of their
types. If the global history of star formation for the composite of all
types of galaxies is determined in this way, cosmological models can then
be tested against the observations of the luminosity density evolution in
the universe. We calculate an evolution of the luminosity density using
the galaxy evolution models of population synthesis developed by Arimoto
\& Yoshii (1987, hereafter AY) and Arimoto, Yoshii, \& Takahara (1992,
hereafter AYT) where the effect of chemical evolution is appropriately
taken into account. Since the population synthesis models of galaxies
are based on the present-day colors of galaxies, the star formation
history to a look-back time corresponding to an average age of stars
in a galaxy is especially well modeled.
These models are therefore reliable especially at low redshifts
of $z<1$ where we constrain the cosmological
parameters from the CFRS data.  

\begin{figure}
  \begin{center}
    \leavevmode\psfig{figure=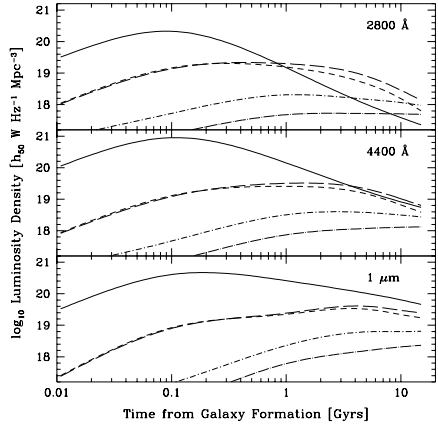,width=6cm}
  \end{center}
\caption{Evolution of the comoving luminosity density at 2800{\AA},
4400{\AA} and 1$\mu$m as a function of time for model galaxies of different
morphological types.  The galactic wind model from AY is used for E/S0
(solid line), and the S1 models from AYT for spiral galaxies of Sab
(short-dashed), Sbc (long-dashed), Scd (dot-short-dashed), and Sdm
(dot-long-dashed).  The luminosity density at 15 Gyrs for the composite
of all types is normalized to
${\cal L}(4400{\rm \AA})=10^{19.296} h_{50}$ W Hz$^{-1}$ Mpc$^{-3}$.
}
\label{fig:l-dens-t}
\end{figure}

\section{Luminosity Density Evolution and Cosmology}
Fig. \ref{fig:l-dens-t} shows the luminosity density
evolution as a function of time for each galaxy type calculated by
the AY and AYT models. For the detail of the calculation, see TYS.
The comparison to the CFRS data is performed in $z<1$, and this
corresponds to $t >$ a few Gyrs in the figure, where $t$ is
the time elapsed from formation of galaxies. (The formation epoch of
$z_F = 5$ is assumed in this figure.)
In short wavebands such as 2800 {\AA}, the luminosity density
is dominated by Sab and Sbc galaxies, while the long waveband of
1 $\mu$m is dominated by E/S0 galaxies. The contribution by late-type
spirals or irregular galaxies is relatively small.

\begin{figure}
  \begin{center}
    \leavevmode\psfig{figure=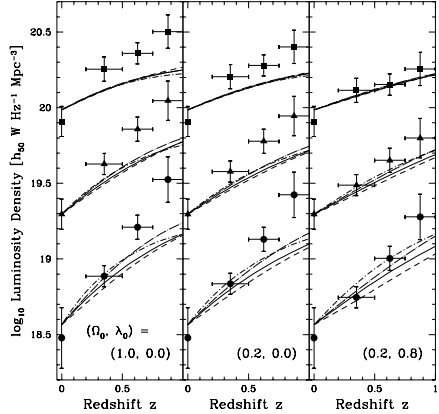,width=6cm}
  \end{center}
\caption{Evolution of the comoving luminosity density as a function of
redshift for the composite of all galaxy types, shown for the
Einstein-de Sitter
universe with $(\Omega_0,\lambda_0)=(1,0)$ ({\it left panel}),
the open universe with (0.2, 0) ({\it middle panel}), and the
$\Lambda$-dominated flat universe with (0.2, 0.8) ({\it right panel}).
The CFRS data points (Lilly et al. 1996),
scaled to $h=0.5$, are shown by circles (2800{\AA}),
triangles (4400{\AA}), and squares (1$\mu$m).  The four theoretical curves
for each of 2800{\AA}, 4400{\AA}, and 1$\mu$m correspond to the evolution
models of S1 (solid line), S2 (dashed), I1 (dot-short-dashed), and I2
(dot-long-dashed) for spiral galaxies.  The galactic wind model for
elliptical galaxies is used in common.  The curves are normalized to
coincide with the 4400 {\AA} data at $z$ = 0.
}
\label{fig:l-dens-z}
\end{figure}

We consider three representative cosmological models: the Einstein-de Sitter
(EdS) universe with
($\Omega_0,\lambda_0)=(1,0)$, an open universe with (0.2,0), and
a $\Lambda$-dominated, flat universe with (0.2, 0.8).  The redshift of
galaxy formation is assumed to be $z_F = 5$.  Different values of $z_F$
hardly change our result as far as we restrict ourselves in $z<1$.
The Hubble constant $h\equiv H_0/$(100 km/s/Mpc) is taken as 0.5, 0.6,
and 0.7 for the three cosmological models in order to give a
reasonable age of galaxies (12.1, 12.4, and 13.6 Gyrs, respectively).
In Fig. \ref{fig:l-dens-z}, we show the comoving luminosity densities
at 2800{\AA}, 4400{\AA}, and 1$\mu$m as a function of $z$ for the EdS
universe ({\it left panel}), the open universe ({\it middle panel}), and
the $\Lambda$-dominated universe ({\it right panel}).  The CFRS data
points are those from the ``luminosity-function-estimated'' values
(Lilly et al. 1996)
for 2800{\AA} (circles), 4400{\AA} (triangles), and 1 $\mu$m (squares).
The four theoretical curves for each waveband correspond to the
variation of the evolution models of spiral galaxies:
S1(solid), S2(dashed), I1(dot-short-dashed), and
I2(dot-long-dashed). These four models are used to estimate the
possible uncertainties in star formation history of spiral galaxies
(see AYT for detail).
It is clear from Fig. \ref{fig:l-dens-z} that
the observed ${\cal L}$-evolution at $z=0-1$ becomes flatter with decreasing
$\Omega_0$ or increasing $\lambda_0$, and eventually falls in agreement
with the $\Lambda$-dominated universe.  In sharp contrast, however, the
observed ${\cal L}$-evolution is too steep to agree with the EdS or open
universe, and such a discrepancy is much more considerable at 1$\mu$m
where elliptical galaxies are dominant and 
uncertainty in star formation history is very small.

For the purpose of quantitative comparison, we have calculated the slope
index of the ${\cal L}$-evolution, $\alpha \equiv d\log{\cal L}/d\log(1+z)$,
averaged over the range of $z=0-1$.
Lilly et al. (1996) estimated the observed
slope for the EdS cosmology as $\alpha = 3.90 \pm 0.75 (2800{\rm \AA})$,
$2.72 \pm 0.5 (4400{\rm \AA})$, and $2.11 \pm 0.5$ (1 $\mu$m).   Using these
observational errors, 
we have performed a $\chi^2$ analysis for $\alpha$ and found that the EdS
and open cosmologies are inconsistent with the data with 99.86\% C.L. and
98.6\% C.L., respectively.  Assuming a flat universe ($\Omega_0+\lambda_0=1$)
with $h=0.6$, the lower limit on $\lambda_0$ is obtained as 0.37 (99\% C.L.)
or 0.53 (95\% C.L.).  Here, one of the galaxy evolution models (S1--I2) is
used that gives the most conservative result in the $\chi^2$ analysis.

\begin{figure}
  \begin{center}
    \leavevmode\psfig{figure=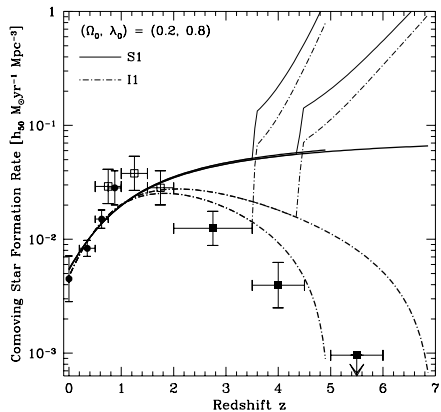,width=6cm}
  \end{center}
\caption{Star formation rate (SFR) in the universe as a function of
redshift.  The $\Lambda$-dominated flat universe with
$(\Omega_0,\lambda_0)=(0.2,0.8)$ is assumed for both predicted and
observed SFRs shown in this figure after scaled to $h=0.5$.   The data
points are from Lilly et al. (1996; filled circles), 
Madau et al. (1997; filled squares)
and Connolly et al. (1997; open squares).
The theoretical curves are based on the S1 (solid)
or I1 (dot-dashed) models for spiral galaxies
and the galactic wind model for elliptical galaxies.
All galaxies are assumed to be formed at $z_F=5$ or 7, with
$h = 0.7$. The SFRs from spiral galaxies only
are shown by thick lines, while those from all types including elliptical
galaxies by thin lines.   The theoretical curves are normalized to agree
with the data at $z<1$.
}
\label{fig:sfr-history}
\end{figure}

\section{Discussion}
Conventional tests on cosmological parameters based on number density
of galaxies, such as galaxy number counts or gravitational lensing
probability in high redshift objects, generally suffer uncertainties
in merger history of galaxies. Most publications of this kind assumed
a constant comoving number density, which has not completely
been confirmed by observations. In contrast, we stress that 
the new test presented
here is based on the relation between evolutionary steepness of
{\it cosmic} star formation rate and
present-day colors of local galaxies, which is essentially determined 
by the stellar evolution theory and the initial mass function.
The luminosity density is relatively insensitive to merger
of galaxies, because
increase in number density of galaxies is compensated by corresponding
decrease in luminosity of each galaxy. Starbursts associated with mergers
might lead to complicated star formation history in a galaxy, but the CFRS has
shown that the luminosity density evolution is smooth at least
when averaged over the universe.

The test on cosmological parameters is performed in $z<1$, where
the uncertainties in both theory and observation are relatively small.
Recent progress in direct observations of star-forming
galaxies at $z\gtilde 2$ however provides important information for
star formation history in the universe.
In Fig. \ref{fig:sfr-history}, assuming the $\Lambda$-dominated universe
with $(\Omega_0,\lambda_0)=(0.2,0.8)$, we plot the observed star formation
rates (SFRs) in the universe which are compiled by Madau (1997) and
Connolly et al. (1997),
together with the SFRs predicted by our model
with $z_F=5$ and 7 assuming $h = 0.7$. Thick lines
show the SFRs from spiral galaxies, and thin lines from all types including
elliptical galaxies.  Since the absolute SFRs are uncertain by a factor of
2--3, we normalize the curves to agree with the data at $z<1$.  Inspection
of this figure clearly indicates that the luminosity density of spiral 
galaxies taking account of gas infall (I1 model)
with $z_F \sim 5$ is consistent with
the observational data within the observational uncertainties.

On the other hand, starbursts in elliptical galaxies, which are
expected from galactic wind models, seem to overpredict the SFR
density in this figure, although an unrealistic assumption of 
one unique formation redshift for all 
ellipticals exaggerates the overproduction.
There are some possibilities to resolve this contradiction:
1) elliptical galaxies are formed at $z_F \gtilde 5$, 2) most of UV
emission from elliptical starbursts is hidden by dust, and 3)
elliptical galaxies are formed without starburst phases, such as
formation from mergers of smaller systems. Maoz (1997) investigated
the first possibility and he found that passively evolving elliptical
galaxies formed at $z \gtilde 5$ are below the detection limits of
the HDF. Meurer et al. (1997) suggest a significant UV extinction by
dust for starburst galaxies at high redshifts and observation of 
the HDF by Infrared Space Observatory also suggests existence of
starbursting ellipticals which could  not be detected in restframe
UV lights (Rowan-Robinson et al. 1997). 
Therefore the overproduction of SFRs in ellipticals
is not so serious at the present stage. Finally, we point out
that the galactic wind model used here ejects 7 \% of its baryon mass
and the amount of metals ejected with this wind is 
sufficient to contaminate the
intergalactic medium as observed in Ly$\alpha$ forests. If the 
escape fraction of UV emission is also as large as 7 \% (for typical
escape fractions, see e.g., Madau \& Shull 1996), the ionizing photons
are enough to reionize the universe. (see TYS for detail.)

This work has been supported in part by the Scientific Research
Fund No. 3730 of the Ministry of Education, Science, and Culture of
Japan.

\end{document}